# Photoinduced strain and polarization switching in barium titanate in the far-infrared spectral range

Maarten Kwaaitaal, Daniel Lourens, Carl S. Davies, and Andrei Kirilyuk*

HFML-FELIX, Radboud University, Toernooiveld 7, 6525 ED Nijmegen, The Netherlands
Radboud University, Institute for Molecules and Materials, 135 Heyendaalseweg, 6525 AJ Nijmegen, The Netherlands

*Contact author: andrei.kirilyuk@ru.nl

**ABSTRACT**

Short mid-infrared laser pulses efficiently facilitate ultrafast manipulation of ferroic order parameters, including full reversal of magnetization or ferroelectric polarization, with the invoked mechanisms relating to the properties of polar phonons in ionic crystals. Much less is known, however, about the behaviour of such order parameters in response to an excitation in the far-infrared range, where phonons are more collective and less polar. Here we investigate transient crystallographic strains and polarization switching in ferroelectric barium titanate ($BaTiO_3$) driven by an excitation in the frequency range of 5-8 THz, or wavelengths of 35-60 μm. We find that switching persists in a large part of this range, but is governed primarily by optical absorption rather than by the longitudinal optical phonons or epsilon-near-zero conditions that dominate in the mid-infrared regime.

## I. INTRODUCTION

In recent years, substantial efforts have sought to acquire efficient and ultrafast control over macroscopic material properties via the optical excitation of microscopic lattice vibrations. Such ionic vibrations, collectively referred to as optical phonons, can be directly stimulated by light pulses tailored to frequencies in the mid- to far-infrared spectral ranges. If the laser pulse is also sufficiently intense, these atomic vibrations can enter a non-linear regime that can temporarily change the crystal structure [1,2]. A series of experimental works have demonstrated that wide-band infrared (IR) laser pulses can induce novel phases in matter [3-5], and control the order parameter in ferroelectric [6-14], magnetic [15-19], and ferroaxial systems [20,21]. The highest phonon amplitudes are achieved when the frequency of the laser pulse is tuned to match the eigenfrequency of an IR-active transverse-optical (TO) phonon mode. One may thus also reasonably expect that light can cause the largest changes when tuned to these frequencies.

Contrary to conventional expectations, experiments using narrow-band IR pulses have revealed that light actually has the largest influence when the frequency is blueshifted relative to the TO phonon mode [9,19,22]. In particular, the observed effects were actually strongest when exciting at the frequency of the longitudinal optical (LO) phonon mode. When these effects were first observed, it was initially hypothesized that the light directly excites these LO modes [19,22], despite the fact that the transversal electric field of the optical pulse (coming at normal incidence) cannot directly couple to such phonons. In spite of the clear inactivity of the LO phonon mode in this scenario, a number of works have repeatedly shown that this mode appears to be critically important in the mid-IR spectral range, allowing magnetization or ferroelectric polarization to be permanently switched. At the same time, it is well-known that there is a regime of strong light-matter interaction when the complex dielectric function ($\varepsilon = \varepsilon_1 + i\varepsilon_2$) approaches zero [23]. Note that the frequency of the LO phonon mode coincides with the frequency at which the real part of the complex dielectric constant ($\varepsilon_1$) is close to zero. Therefore, it has recently been suggested that the LO mode does not serve as the main driver for switching at these wavelengths, but rather the ENZ (epsilon-near-zero) condition is actually responsible [9].

Thus far, the relevance of LO phonons and the ENZ condition is empirically based on experimental results obtained by exciting ferrimagnetic, antiferromagnetic or ferroelectric samples with mid-IR laser pulses having wavelengths below 30 µm (frequencies above 10 THz). There are, however, additional optical phonon modes in the far-infrared spectral range, whose influence under laser excitation have not yet been investigated. A natural question then emerges - does the LO phonon mode, or equivalently the epsilon-near-zero regime, keep its importance when going to the far-infrared spectral range?

To explore this question, we consider the prototypical and extensively-studied ferroelectric barium titanate, which displays all-optical switching of polarization in the mid-infrared range [9] and whose phonon modes in the far-infrared range are well known [24,25]. While the characteristic optical phonons of BaTiO$_3$ in the far-infrared region have much stronger oscillator strengths compared to their siblings in the mid-infrared range, the far-infrared phonon modes are substantially overdamped (see Fig. 1). It is therefore interesting to assess whether these differences will affect the spectral behaviour of light-induced changes near these spectral positions.

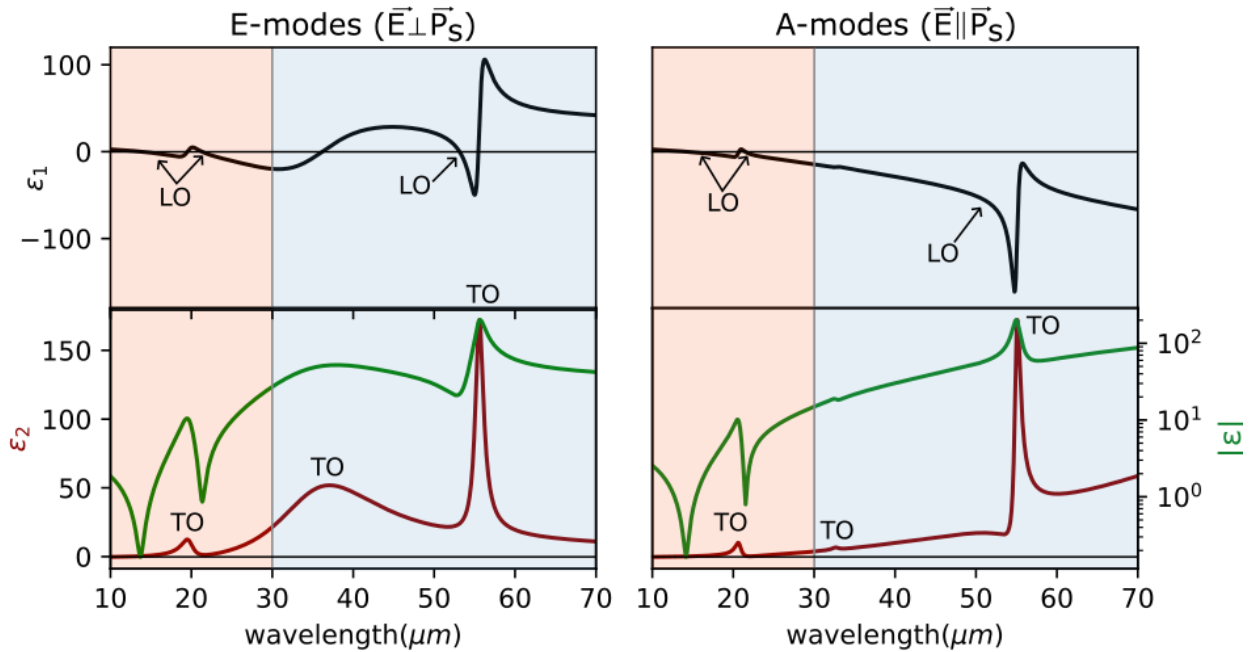

*Figure 1: Real ($\varepsilon_1$) and imaginary ($\varepsilon_2$) parts of the dielectric function of BaTiO$_3$ in the mid- and far-infrared (IR) ranges, as well as absolute values of $\varepsilon$ [24]. The response in the left and right panels is obtained when the light's electric field E is perpendicular and parallel to the ferroelectric polarization $P_s$, coupling to E- and A-phonon modes, respectively. The spectral positions of the longitudinal optical (LO) and transverse optical (TO) phonon modes are indicated. Note that the optical constants are much smaller in the mid IR region compared to the far IR region.*

In this work, we study how far-IR light pulses induce strain and polarization switching in ferroelectric BaTiO$_3$. To pump this material, we use a free-electron laser to obtain narrow-band tunable IR pulses with central frequencies ranging between 5 THz and 8 THz (wavelength 35-60 µm). The resulting strain and polarization switching are dynamically tracked via birefringence and second-harmonic-generation microscopy. We find that, in the far-IR spectral range, switching can still be driven in specific wavelength regions. However, the optical absorption properties appear to become much more relevant for the switching, compared to the strongly resonant LO-phonon / ENZ condition which was important in the mid-infrared range.

## II. METHODS

### A. Material and imaging techniques

In the tetragonal ferroelectric phase of BaTiO$_3$ (space group *P4mm*), the zone-centre optical phonons split according to the irreducible representations $3A_1 + B_1 + 4E$. The $A_1$ modes involve atomic displacements along the tetragonal *c*-axis, while the doubly-degenerate *E* modes correspond to ionic vibrations in the *ab* plane. Both the $A_1$ and *E* modes are Raman- and IR-active, whereas the $B_1$ mode is Raman-active only. Experimentally and from lattice-dynamical calculations [25], the typical TO frequencies at room temperature are $A_1(TO_1) \approx 180$ cm$^{-1}$, $A_1(TO_2) \approx 270$ cm$^{-1}$, and $A_1(TO_3) \approx 520$ cm$^{-1}$; $E(TO_1) \approx 40$ cm$^{-1}$, $E(TO_2) \approx 180$ cm$^{-1}$, $E(TO_3) \approx 305$ cm$^{-1}$, and $E(TO_4) \approx 490$ cm$^{-1}$; and $B_1 \approx 305$ cm$^{-1}$. Because the $A_1$ and *E* modes are polar, they exhibit LO-TO splitting due to long-range Coulomb interactions, with the corresponding longitudinal optical frequencies being $A_1(LO_1) \approx 180$ cm$^{-1}$, $A_1(LO_2) \approx 470$ cm$^{-1}$, and $A_1(LO_3) \approx 720$ cm$^{-1}$; and $E(LO_1) \approx 180$ cm$^{-1}$, $E(LO_2) \approx 305$ cm$^{-1}$, $E(LO_3) \approx 470$ cm$^{-1}$, and $E(LO_4) \approx 720$ cm$^{-1}$.

By virtue of its tetragonal phase, barium titanate has a total of six possible equilibrated polarization directions. These polarization orientations are perpendicular to each other, leading to orientations approximately parallel to the $\pm x, \pm y, \pm z$ direction of a Cartesian coordinate system. At room temperature, the perovskite has a tetragonal crystal structure with a difference in optical properties along the ferroelectric polarization axis and perpendicular to it. One of these properties is birefringence, with barium titanate being a uniaxial crystal in which the refractive index along the spontaneous polarization differs. As a consequence of this birefringence, barium titanate can modify the ellipticity of polarized light traveling through the crystal, which can be measured by a polarization-sensitive microscope. The changes in the light's ellipticity depend on the domain orientation, allowing us to distinguish domains where the polarization direction differs by 90° [26]. Furthermore, the magnitude of the birefringence depends also on the response of external stimuli, such as laser-induced strain and heating [27], allowing us to additionally observe these effects.

For our experiments, we use a double-side-polished, 0.5-mm-thick BaTiO$_3$ crystal. To visualize strain and 90° domains, we use a polarization microscope with light of wavelength 520 nm for illumination. The latter was provided by one of two sources, depending on the timescale that was explored. The first light source is a continuous-wave diode laser (Integrated Optics), with time-resolution provided by electronically delaying the CCD camera (exposure time 27 µs). To achieve better time-resolution, we also replaced the CW-laser with a 5-ns-long pulse delivered by a frequency-doubled Nd:YAG laser, which thus provides an experimental time resolution of 5 ns.

To distinguish domains where the ferroelectric polarization differs by 180°, we instead use second harmonic generation (SHG) microscopy. To provide illumination, we use 150-fs-long pulses of wavelength 1040 nm delivered by an amplified Yb-doped fiber laser. The laser pulses have a repetition rate of 100 MHz and energies up to 90 nJ. A detailed description of the exact setup used for these SHG domain imaging experiments can be found in Ref. [28].

**B. Excitation with far-infrared light**

For the experiments, we used tuneable radiation between 35 μm and 65 μm from the FELIX free-electron laser [29]. The laser produces picosecond-long pulses with a narrow bandwidth (rms typically 0.5-1%). Depending on the experiment, these pulses have a repetition rate of 50 MHz, coming within a single 10-μs-long burst ("macropulse"). The total energy of such a macropulse is wavelength-dependent, varying between 1 mJ and 3.5 mJ in the 50 MHz mode. The laser is focussed on the sample at normal incidence to a spot with diameter ≈300 μm.

## III. RESULTS AND DISCUSSION

**A. Photoinduced switching of 90° domains**

We start our experiments by studying the process of 90°-domain switching. Typical images of the sample taken before and after excitation by a single infrared macropulse of wavelength 42 μm are shown in Fig. 2. The switched domains appear as thin stripes, which is the common shape for domains in barium titanate. In addition, we see a circular change in contrast change that spans the entire irradiated area. This is caused by a temperature increase resulting from the absorption of the laser light in combination with a temperature-dependent birefringence of the sample [26]. This easily observable effect thus gives us information both about the size of the laser spot and about the temperature increase caused by the excitation.

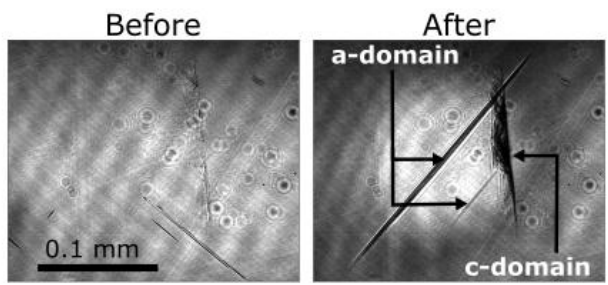

*Figure 2: Polarization microscope images taken before and after laser irradiation at the wavelength 42 μm. After the laser pulse, there appear diagonal stripes, which are newly formed a-domains and c-domains as indicated. A faint ring of contrast is also observed, which stems from thermal modifications of the crystal's birefringence [27]*

The switched domains are most often transient in nature. To therefore investigate their dynamics, we excited the sample with a single macropulse of wavelength 42 μm, and probed the changes with a 5-ns laser pulse. An electronic delay generator allows the time delay between the pump and probe excitations to be adjusted arbitrarily, across a timescale of several hundreds of microseconds. The cumulative size of the created domains as a function of time is shown in Fig. 3. The appearance and growth of the new domains take place throughout the macropulse and a few microseconds afterwards. The growth is followed by a stable period of time lasting ≈50 μs, before the switched domains start to gradually shrink and vanish on a timescale of hundreds of microseconds.

Investigating the influence of the pump's light polarization relative to the ferroelectric polarization could provide deeper insight into the origin of the switching process as it is an important parameter in several theoretical studies [11,30]. In our setup, the linearly-polarized IR pulse arrives at normal incidence to the surface. Because the initial ferroelectric polarization was oriented along the surface plane, we could therefore align the optical polarization at varying angle relative to the initial ferroelectric polarization. Figure 4(A) shows how the size of the switched domain depends on this relative angle, with fixed incident pulse energy. Overall, we observe the largest switching when the IR light's electric field lies collinear with the ferroelectric polarization, causing the ferroelectric polarization to switch 90° both within or normal to the crystal surface. Moreover, upon rotating the optical polarization away from the ferroelectric polarization, the amount of switching diminishes. For the selected pulse energy, switching was completely absent when the polarization of the IR excitation was oriented at angles of 60° or more relative to the ferroelectric polarization. Note that increasing the pulse energy leads to an overall increase in switching probability, enabling switching also when the light polarization is perpendicular to the ferroelectric moment.

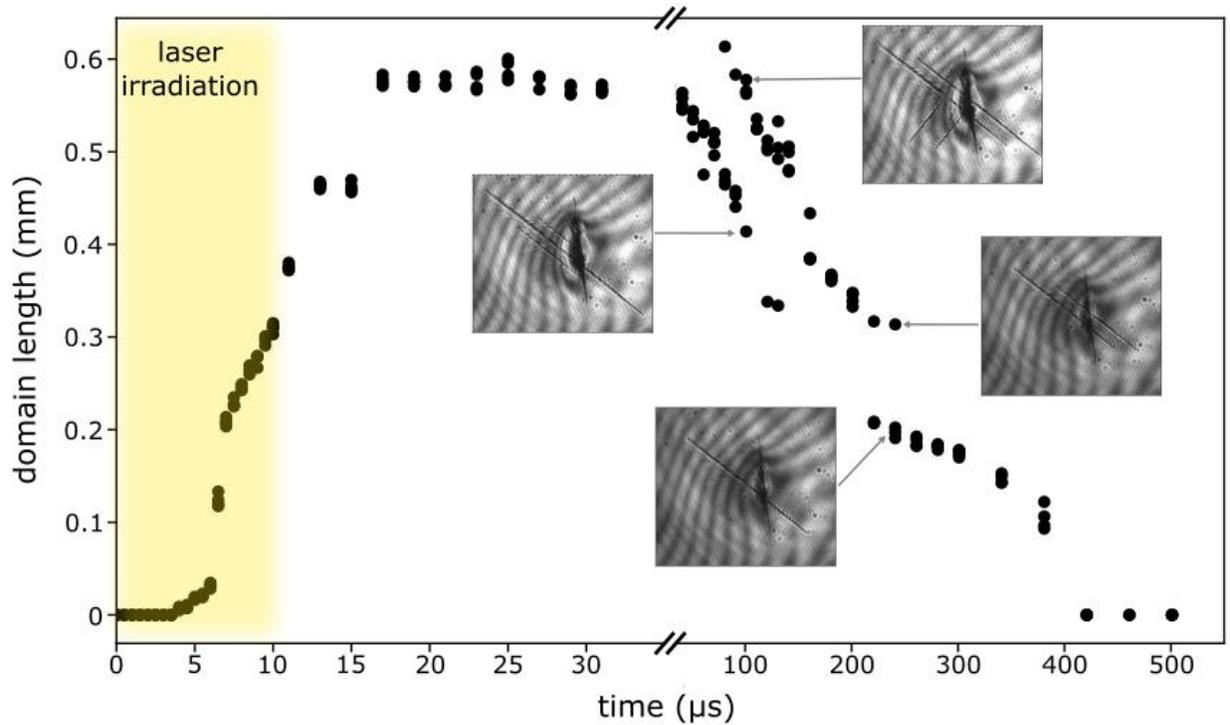

*Figure 3: Domain dynamics after excitation by a 10 µs-long macropulse excitation of wavelength 42 µm. We measured the length of the in-plane domains (Diagonal stripes). The data contains two possible trends for the decay of the newly formed domains, corresponding to some randomness in the formed domain pattern. To further demonstrate this, we twice provide a pair of images (width of 0.25 mm) obtained at the same time delay.*

The angular dependence shown in Fig. 4(A) could stem from two possible sources. On one hand, the underlying mechanism for switching could be more efficient when the impinging light has optical polarization collinear with the ferroelectric polarization. On the other hand, anisotropy in the optical properties of $BaTiO_3$ could also contribute to the observed polarization dependence. In particular, at the spectral line about 42 µm, the reflectivity of barium titanate is significantly diminished when light is polarized along the ferroelectric polarization compared to the case of perpendicular orientation [24]. The red line in Fig. 4(A) shows the fraction of light penetrating in the medium, which is given by 1-$R$ (where $R$ is the reflectivity). We clearly observe that the switching decreases together with the amount of pulse energy that penetrates the sample. It is therefore likely that the angular dependence shown in Fig. 4(A) mainly originates from the reflectivity changes.

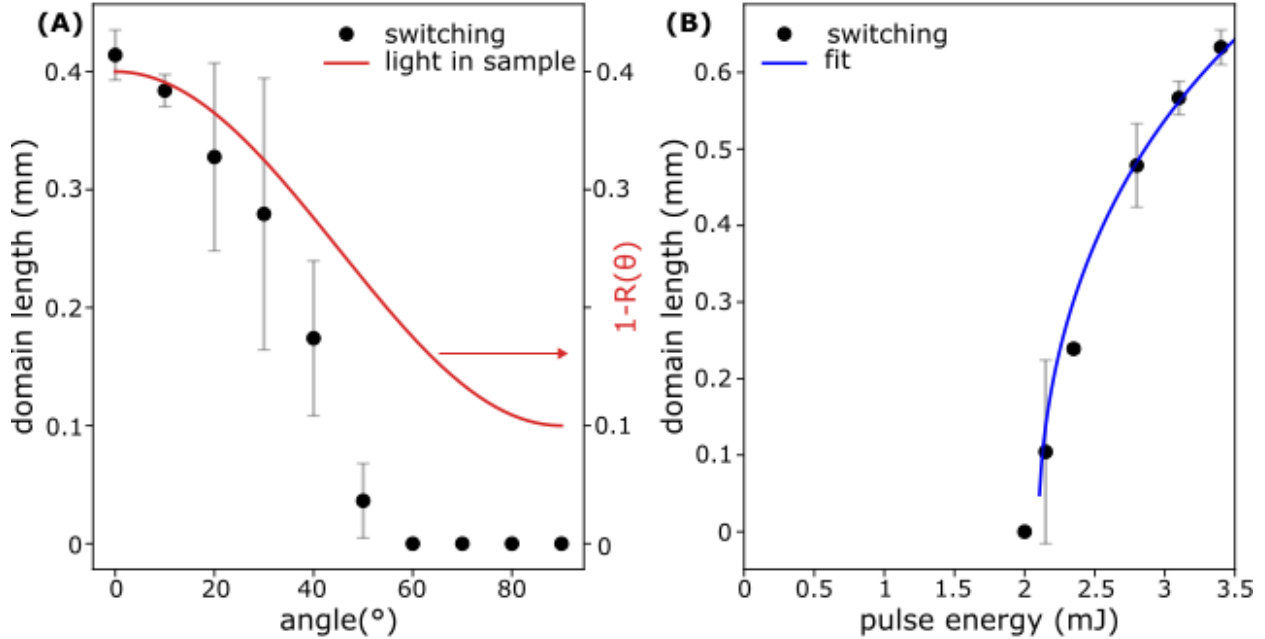

Figure 4(A) Black dots: size of switched domains as a function of the pump's polarization. Additionally, the red line shows the fraction of light that penetrates the sample which is given by 1-$R$ (where $R$ is the reflectivity). At 0°, the optical polarization is aligned to the ferroelectric polarization. (B) Switching as function of pulse energy when the IR light is polarized along the ferroelectric axis. Both measurements are performed with a wavelength of 42 $\mu$m.

The switching process has further been investigated at the spectral line of 39 $\mu$m, which correlates with no particular optical phonons. Here, we vary the pump pulse energy when the light is polarized along the ferroelectric polarization. The results of this measurement are shown in Fig. 4(B). We clearly observe that a certain threshold amount of energy is required before switching can be observed. When above this threshold, the domain size grows with pulse energy, albeit the slope of this growth becomes smaller at higher energies. Such behaviour is commonly associated with threshold effects [31] and has previously been used to describe the energy dependence of optical switching of magnetic domains [32]. For processes with a Gaussian laser spot (of radius $a$) that require a threshold energy c ($E_{thresh}$), we obtain the following condition:

$$E_0 \cdot e^{-\frac{r^2}{2a^2}} > E_{thresh} \tag{1}$$

For a switching process, the domain size would therefore be related to the critical radius $r$ that satisfies the above condition, given by

$$r < a\sqrt{2 \ln\left(\frac{E_0}{E_{thresh}}\right)} \tag{2}$$

This relation is fitted to the domain size versus pulse energy data in Fig. 4(B), yielding a threshold fluence of 13 J/cm². The fit clearly matches the experimental data rather well.

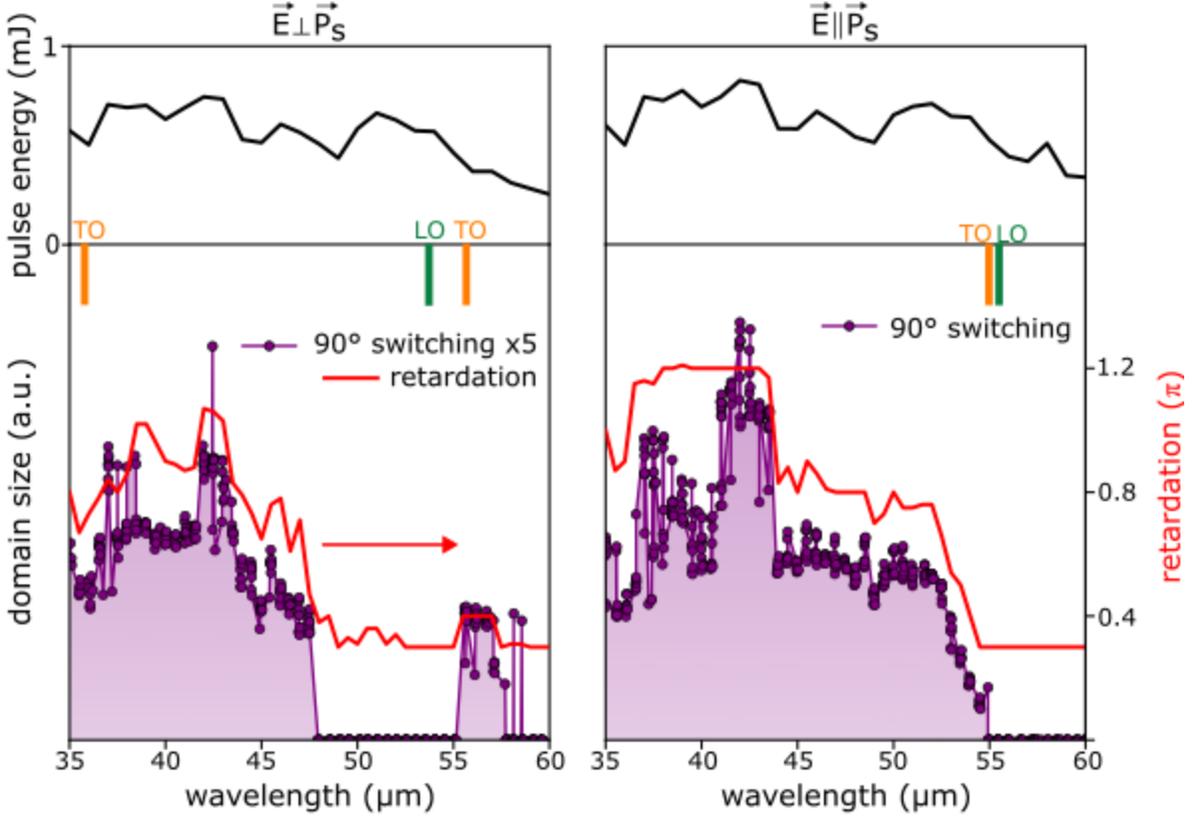

*Figure 5: Wavelength dependence of 90° switching and heat induced retardation for both polarization configurations. The top panel shows the pulse energy as a function of wavelength during the measurement. The bottom panel shows the size of switched domains (purple graph) and the retardation (red line). Additionally, the position of the TO and LO phonon modes are shown as green and orange lines.*

Next, we explored the influence of the excitation wavelength on the switching. We focus our attention to the wavelength range of 35-60 µm, and distinguish between the cases where the optical polarization is parallel and perpendicular to the ferroelectric polarization. The results, presented in Fig. 5, show that switching can be successfully achieved across this spectral range. We also overlay, in Fig. 5, the position of the phonon modes of $BaTiO_3$ in this range. Surprisingly, there appears to be no correlation between the phonon modes and the switching. If the phonon modes do not therefore drive the switching, how can the switching emerge?

To answer this question, we also measured the retardation change at the centre of the laser spot, which is visible in the polarization change as concentric rings and can be found by integrating the birefringence over the thickness of the sample. The change is caused by a thermally-induced reduction of the birefringence, which falls to zero upon reaching the Curie temperature. It thus gives an estimate of the total heat locally generated in the sample by the pump pulse. The switching has a good spectral match with that of the retardation, suggesting that laser-induced heating plays an important role in the switching process. In fact, several studies [33,34] have already explored how laser heating can switch domains in lithium niobate. In those cases, the heating served to raise the sample temperature towards the Curie temperature, which in turn reduces the coercive field. In addition, the temperature gradient due to non-uniform laser heating creates an inhomogeneous electric field through the Seebeck effect, which can therefore switch the ferroelectric polarization. It is possible that a similar mechanism is responsible for the switching observed here.

It is interesting to compare these results with previous experiments on laser-induced switching in ferroelectrics. First of all, the experiments with visible light require very narrowly focused beams to create large enough temperature gradients and polarization changes are parallel to the sample surface. Linear absorption is very low in the visible region for ferroelectric materials such as barium titanate and lithium niobate. In contrast, at mid-infrared frequencies, absorption is much higher. Therefore, in our experiments, the excitation spot is relatively large (200 µm) and the temperature gradient is mainly perpendicular to the surface. Consequently, the switched domains in our experiments mainly have their polarization perpendicular to the surface. It should be noted that in-plane switching is also possible for cases where switching efficiency is highest. Nevertheless, out of plane switching dominates.

Closely related to these observations are our previous results on ferroelectric switching [9], obtained when the pump excitation was tuned to the highest-frequency phonon modes in the mid-IR range (excitation wavelengths of 8-25 µm). In these cases, switching manifested at the LO phonon frequencies, and was predominantly in-plane. Furthermore, there is a spectral mismatch between the retardation change and the switching at the high-frequency region [9,27]. Thus, we conclude that the switching behaves differently depending on whether the excitation targeted the higher- or lower-frequency phonon modes. Although laser heating is unavoidably present in both regions, it appears that heating is the dominant mechanism at lower frequencies, but that there is another dominant cause for switching at the higher frequencies.

**B. 180° switching**

Optical second harmonic generation microscopy is employed to monitor the dynamics and spectral behaviour of 180° domains [28]. A typical example of the induced switching is shown in Fig. 6 (A,B), revealing a characteristic spatial pattern that is quite peculiar. In particular, despite the impinging laser having a Gaussian spatial distribution of electric field, the switched domains do not appear at the centre (where laser amplitude is highest) but rather at the sides of the laser spot. This distribution of domains has previously been explained by the generation of laser-induced piezoelectric displacement fields [9].

In principle, strain couples primarily to ferroelastic degrees of freedom and therefore one would expect it to be involved in 90° domain switching, which have a different spontaneous strain. On the contrary, the difference between the 180° domains is in the ferroelectric nature only and thus the effect of strain is less obvious. Nevertheless, strain can influence the ferroelectric polarization via the piezoelectric effect, which can enhance or reduce the ferroelectric polarization depending on whether the strain is compressive or tensile. The strain profile introduced by a laser pulse is described in more detail in Section 5 and Ref. [9]. It has been shown that this profile enhances the initial ferroelectric polarization at the centre of the laser pulse and reduces it at the sides, making switching easier at the exact locations where we observe the switched domains, while switching would be more challenging at the centre of the spot with the highest laser intensity. We draw attention to the point that gradient-based mechanisms, such as the flexoelectric or thermoelectric effects, could not explain this domain pattern since switching appears on both flanks of the irradiated area. These mechanisms inherently produce regions with opposite gradient signs at the irradiated area (positive and negative), which would lead to switching at only one flank rather than both.

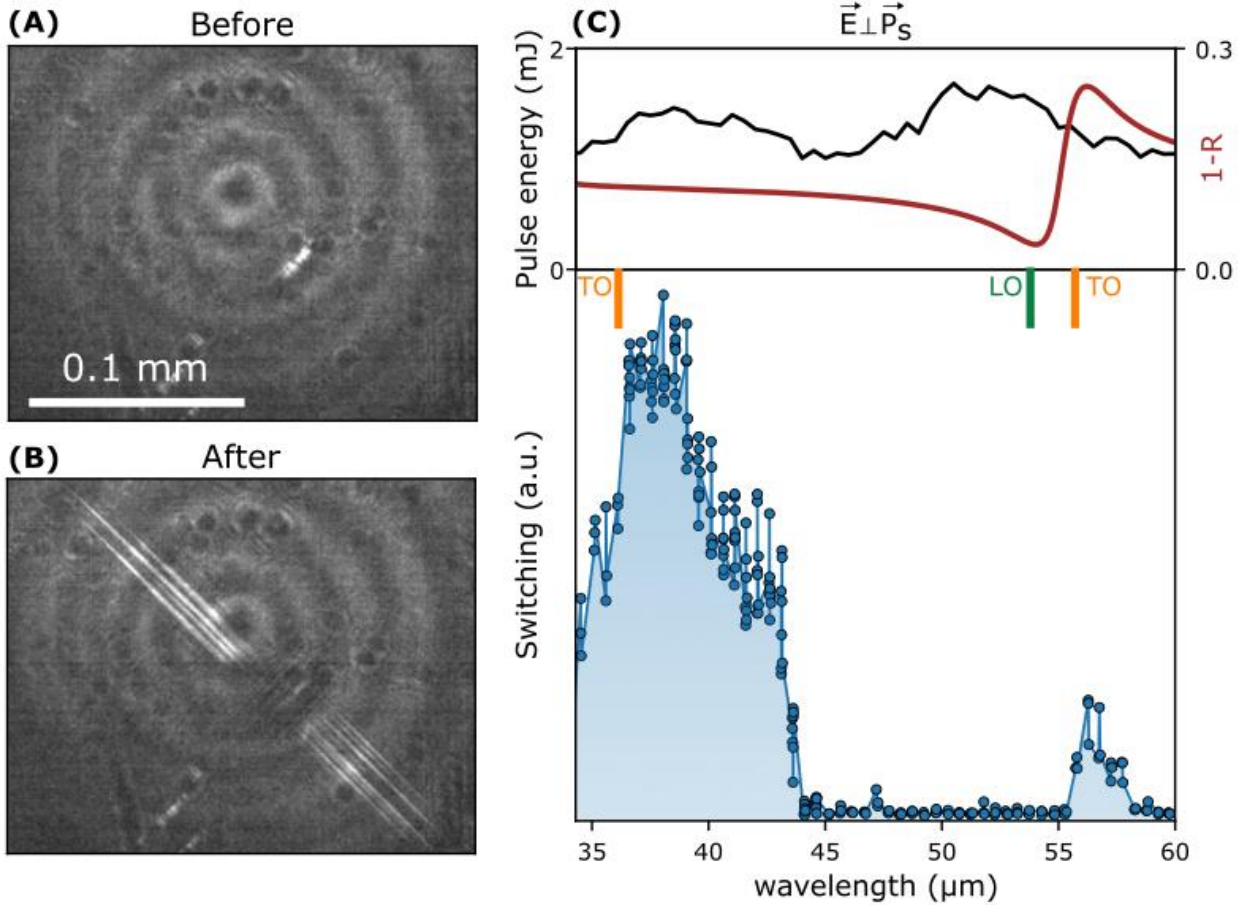

*Figure 6: Left image: Second harmonic generation microscope images taken before and after laser irradiation (of wavelength 40 µm). The bright areas are the regions correspond to 180° domains. The laser spot is centered between the two switched domain areas in the after image. Right column, bottom panel: wavelength dependence of 180° switching. Switching occurs mainly between 35-45 µm and there is a small peak just above 55 µm. This last peak matches a decrease in the reflectivity (see top panel)*

By taking images before and after optical irradiation, we assess the size of the 180° switched domains as a function of wavelength (Fig. 6, right column). The spectral dependence broadly resembles that of the 90° domain switching. There is a large switching window at the smaller wavelengths (34 µm ≤ λ ≤ 44 µm), above which switching vanishes. At λ ≈ 57 µm, there is again some clear emergence of switching. This small peak can be explained by a rapid variation of the reflectivity in the vicinity of this wavelength. At λ ≈ 57 µm, the reflectivity is minimized, which increases the fraction of light penetrating the sample (measured by the quantity '1-$R$', which is plotted in the top panel of Fig. 6).

**C. Strain**

Optical excitation by a focused laser beam generally leads to a local deformation of the material, the shape of which is determined by the excitation profile. In practice, our beam profile has a radially-symmetric Gaussian profile. This leads to a uniform strain profile that has been observed and described previously in Ref. [27] and visualized for our case in Fig. 7(A). The resulting shear-strain ($S_{xy}$) can be observed using polarization microscopy due to the elasto-optical effect. The shear-strain, originating from a radially symmetric excitation profile, is expected to consist of four lobes, as shown in Fig. 7(B). The corresponding contrast changes as experimentally observed in the polarization microscope are shown in Fig. 7(C) and matches this pattern well.

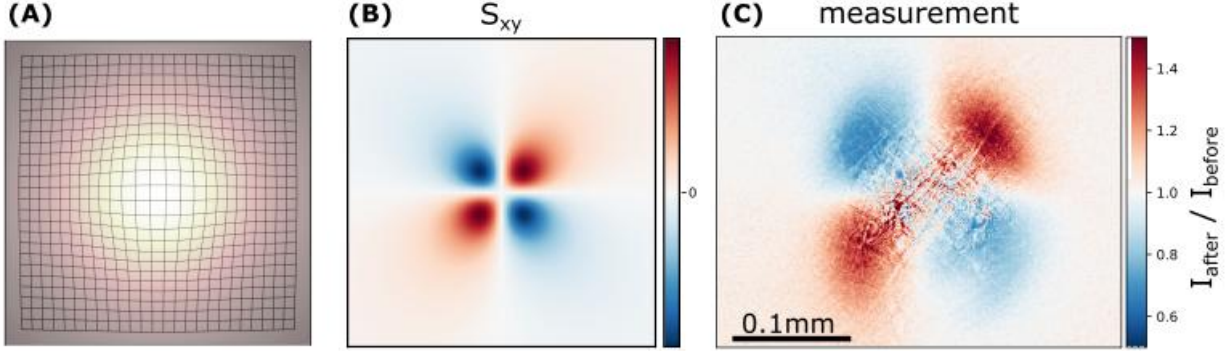

*Figure 7: (A) Two-dimensional-visualization of lattice expansion due to optical irradiation, following the Gaussian intensity profile of the laser spot. This expansion is accompanied by a strain profile. (B) Theoretical calculation of shear strain stemming from the expansion. (C) Experimental visualization of the shear strain in the polarization microscope. The contrast shown is a difference ratio between the probe intensity taken before and after laser irradiation.*

The question now is how we can use the images acquired through polarization microscopy to quantify the amount of strain induced by the laser excitation. In general, the optical intensity of the probe that passes through the polarization microscope can be calculated with Jones calculus. In the limit of the polarizer and analyzer being crossed, the obtained expression simplifies to

$$I = I_0 \cdot sin^2(2\alpha) \cdot sin^2\left(\frac{\Delta\varphi}{2}\right), \qquad (3)$$

with $\Delta\varphi$ being the retardation in the material and $\alpha$ the angle that the optical axis of the crystal makes with the transmission axis of the polarizers. Shear strain can modify the optical axis of a material and thus changing the angle that this axis makes with respect to the polarizers. The strain perturbation of the angle is relative small, hence we will write as the combination of an initial angle $\alpha_0$ and the strain induced change $\alpha_s$: $\alpha = \alpha_0 + \alpha_s$, where the strain part is given by [27]:

$$\alpha_s = 2 \frac{B_{xy}}{B_{yy} - B_{xx}}. \qquad (4)$$

Here the impermeability $B_{ij}$ is changed by the strain ($\Delta B_{ij} = p_{ijkl} S_{kl}$), where $p_{ijkl}$ are the elasto-optical coefficients. Furthermore, to obtain contrast, we rotate the sample such that the unstrained crystal axis is almost parallel to the polarizers. i.e. $\alpha_0, \alpha_s \ll 1$ and apply a Taylor expansion:

$$sin^2 \alpha \approx sin^2 \alpha_0 + 2 sin(\alpha_0) cos(\alpha_0) (\alpha_s - \alpha_0) + (cos^2(\alpha_0) + sin^2(\alpha_0))(\alpha_s - \alpha_0)^2 \qquad (5)$$

Practically speaking, the best contrast was achieved when rotating the sample under a small angle to allow more light through and slightly misaligning the polarizer and analyzer angle such that they are not perfectly perpendicular to each other. In this case one observes two bright and two dark lobes. The observation of the strain pattern through the polarization microscope is shown in Fig. 7(B,C).

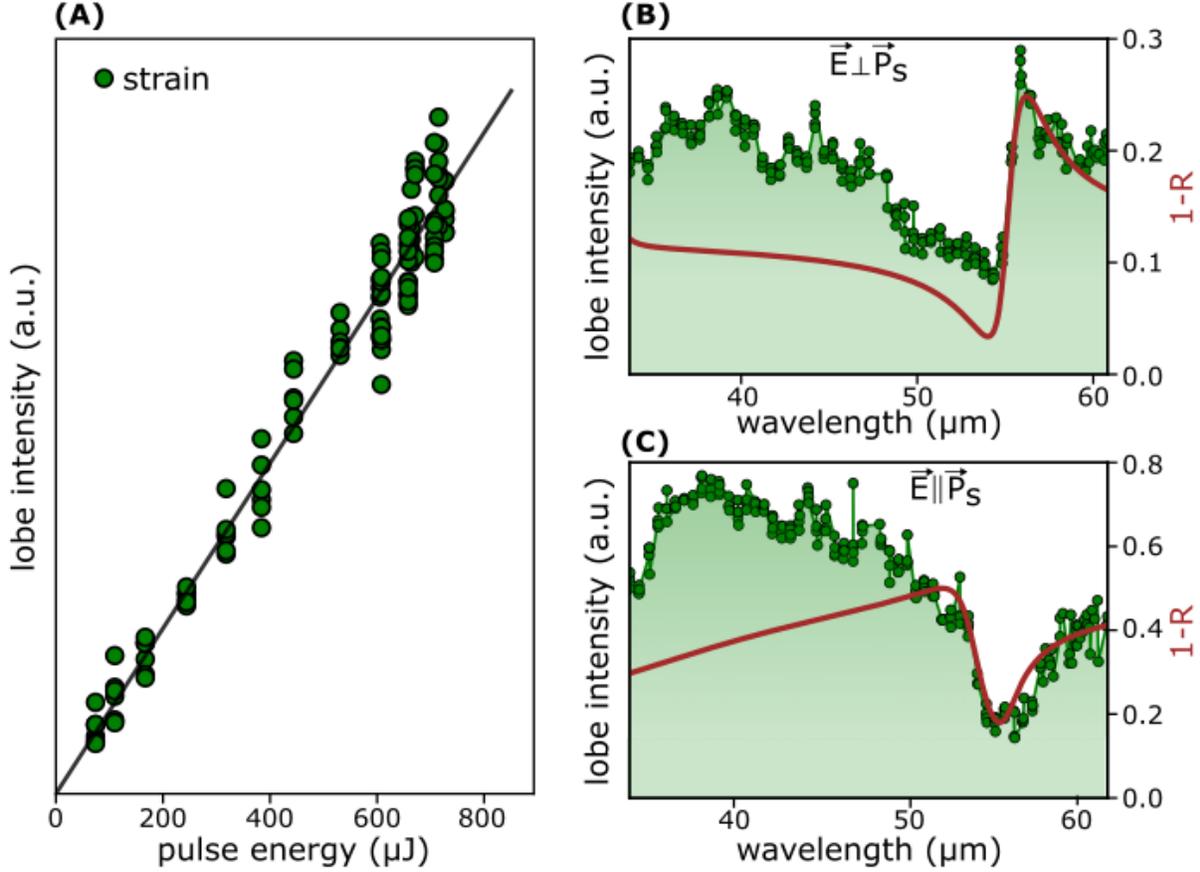

*Figure 8: (A) Energy dependence of the lobe intensity change at the pump wavelength of 37 µm, with an accompanying linear fit. (B,C) Wavelength dependence of the change in the lobe intensity, normalized by the wavelength-dependent pulse energy. The strain measurement (green graph) shows a clear spectral resembles with 1-R (brown line), which is the amount of absorbed light by the sample.*

Using the expression above, we measured the lobe intensity change as a function of the excitation pulse energy (Fig. 8(A)), showing a linear behaviour. We have also verified that a similar linear dependence between the pulse energy and the intensity change of the lobes is observed at all wavelengths in the used range. From this, in combination with the first order Taylor expansion (Eq. (5)), we conclude that strain scales linearly with the energy of the infrared laser pulse.

The linear relation between pulse energy, strain and lobe intensity actually provides an excellent diagnostic for investigating the spectral dependence of the amount of photo-induced strain (Fig. 8(B,C)). Upon changing the excitation wavelength of the laser pulse, the pulse energy also changes. Unlike the switching that shows a threshold behaviour, the generated strain can be straightforwardly normalized for the pulse energy. The strength of the photoinduced strain is relatively constant over the spectrum with an exception of around $\lambda \approx 55$ µm, where there is a TO phonon mode. There is however no clear resonance visible around this phonon mode, implying that the phonon excitation is not the main reason for the generation of the strain. Upon closer inspection, the strain follows very well the wavelength-dependent changes in the amount of light that penetrates the sample (proportional to the quantity '1-R').

**IV. CONCLUSION**

We have demonstrated that all-optical switching of the ferroelectric polarization in BaTiO$_3$ is not only possible in the mid-infrared spectral range [9], but also in the far-infrared. The spectral behaviour of the switching in these two regimes, however, is distinctly different. In the mid-infrared range, the spectrum of switching follows the frequencies of LO phonon modes, which correlate with both the real and imaginary parts of the dielectric function being close to zero. As a result, the absolute value of the dielectric function reveals deep minima, as illustrated in Fig. 1. In Ref. [9], the switching is therefore explained in terms of optical enhancement stemming from the epsilon-near-zero condition that drastically increase the light-matter interaction. In the far-infrared range in contrast, the optical phonons are overdamped. As a result, only the real part of the dielectric function goes close to zero at the LO phonon frequency, whereas the imaginary part remains large, and so the absolute value of ε stays large. Note that from the point of view of ENZ-driven effects, the absolute value of ε should be smaller than 2 [23], which is only achieved in the mid-IR range (see Fig. 1). This leads to a different spectral dependence without clear resonances around this phonon mode.

In the far-infrared range, the spectra of switching and strain appears to scale inversely with the reflectivity of the material. When the reflectivity is low, more light penetrates the sample, and thus absorption and photoinduced heating is enhanced. This appears to play a more dominant role in the process of switching in the far-infrared range. This conclusion is supported by the spectral correlation between absorption and 90° and 180° switching. Nevertheless, it is difficult to explain the complex distributions of switching patterns in terms of thermal effects only. Several non-thermal pathways could also contribute to the switching [8,10-14,30]. At present, our use of a pumping macropulse inescapably leads to cumulative heating, built up over the ≈500 pulses within the 10-µs-long burst. Further research should aim to clarify whether a single laser pulse in the far-infrared spectral range can lead to similar switching.

## V. ACKNOWLEDGEMENTS

The authors thank all technical staff of the FELIX facility for technical support. D.G.L. acknowledges funding by the Max Planck–Radboud University Center for Infrared Free Electron Laser Spectroscopy. C.S.D. acknowledges support from the European Research Council ERC Grant Agreement No. 101115234 (HandShake), and A.K. acknowledges support from the European Research Council ERC Grant Agreement No. 101141740 (INTERPHON).